\documentclass[aps,prl,reprint,superscriptaddress]{revtex4-1}
\usepackage[utf8]{inputenc}
\usepackage[english]{babel}
\usepackage{amsmath}
\usepackage{amsfonts}
\usepackage{amssymb}
\usepackage{graphicx}
\usepackage{epstopdf}
\usepackage{url}
\usepackage[usenames]{color}
\begin{document}
\author{P.N.~Skirdkov}
\author{K.A.~Zvezdin}
\affiliation{A.M.~Prokhorov General Physics Institute, Russian Academy of Sciences, Vavilova 38, 119991 Moscow, Russia}
\affiliation{Moscow Institute of Physics and Technology, Institutskiy per. 9, 141700 Dolgoprudny, Russia}
\affiliation{Russian Quantum Center, Novaya St. 100, 143025 Skolkovo, Moscow Region, Russia}
\author{A.D.~Belanovsky}
\affiliation{A.M.~Prokhorov General Physics Institute, Russian Academy of Sciences, Vavilova 38, 119991 Moscow, Russia}
\affiliation{Moscow Institute of Physics and Technology, Institutskiy per. 9, 141700 Dolgoprudny, Russia}
\author{J.M.~George}
\affiliation{Unité Mixte de Physique CNRS/Thales, 1 avenue Augustin Fresnel, 91767 Palaiseau, France, and Université Paris Sud, 91405 Orsay, France}
\author{J.C.~Wu}
\affiliation{National Changhua University of Education, 1, Jin-De Road, Changhua 500, Taiwan}
\author{V.~Cros}
\affiliation{Unité Mixte de Physique CNRS/Thales, 1 avenue Augustin Fresnel, 91767 Palaiseau, France, and Université Paris Sud, 91405 Orsay, France}
\author{A.K.~Zvezdin}
\affiliation{A.M.~Prokhorov General Physics Institute, Russian Academy of Sciences, Vavilova 38, 119991 Moscow, Russia}
\affiliation{Moscow Institute of Physics and Technology, Institutskiy per. 9, 141700 Dolgoprudny, Russia}
\affiliation{Russian Quantum Center, Novaya St. 100, 143025 Skolkovo, Moscow Region, Russia}
\title{Large amplitude vortex gyration in Permalloy/Bi$_2$Se$_3$-like heterostructures}
\begin{abstract}
We consider the excitation of large amplitude gyrotropic vortex core precession in a Permalloy nanodisk by the torques originating from the in-plane microwave current flowing along the interface of the Permalloy/Bi$_2$Se$_3$ heterostructures, in which the huge charge-to-spin conversion ratio is observed \cite{Mellnik-2014}. We consider analytically and by micromagnetic modelling the dependence of this excitation on the frequency and magnitude of the microwave current. The analogies of the vortex dynamics and the Landau phase transitions theory is demonstrated. These findings open the possibility to excite gyrotropic vortex motion with the current densities far lower than by any other means.
\end{abstract}
\maketitle
Magnetic vortices are very interesting nanoscale magnetic states which attracted research interest from both fundamental and applied perspectives \cite{Guslienko-2001, Guslienko-2002, Novosad-2002, Ivanov-2005, Guslienko-2005, Shibata-2006, Guslienko-2006, Ivanov-2007, Pribiag-2007, Kruger-2007, Bolte-2008, Dussaux-2010, Sugimoto-2011, Drews-2012}. Magnetic vortices are formed by in-plane magnetization that curls clockwise or counter-clockwise around the small volume of out of plane magnetization in the center which is called the vortex core. This magnetization distribution in nanopillars \cite{Shinjo-2000, Wachowiak-2002} represents an energy favourable state for a wide range of geometries \cite{Cowburn-1999, Guslienko-2004}. 
\par
The increased attention to magnetic vortices is largely related to their excited states, which can be maintained by an external stimuli (field, current etc.). Among the dynamic modes of magnetic vortices in nanopillars, the gyrotropic mode is of the greatest interest, because it is the lowest energy collective excitation \cite{Zaspel-2005, Guslienko-2006, Ivanov-2007}. It is represented by the circular (or gyrotropic) motion of the vortex core, accompanied by corresponding high-amplitude oscillations of the mean magnetization of the whole nanopillar. There are promising reports on the possibility to build vortex-based non-volatile memories \cite{Bohlens-2008, Pigeau-2010}, Spin Transfer Nano-Oscillators (STNOs) \cite{Pribiag-2007, Dussaux-2010}, spin diode frequency sensors \cite{Jenkins-2014}, and even more sophisticated systems based on collective vortex dynamics \cite{Jung-2011}. 
\par
The key issue in this context is the manner of the vortex gyrotropic mode excitation. Initially, it was proposed to excite vortices by an external microwave magnetic field (see for instance Refs.~\cite{Waeyenberge-2006, Chou-2007, Kruger-2007}). An alternative approach is to use dc spin-polarized current to excite the gyrotropic motion of the vortex (see for instance Refs.~\cite{Pribiag-2007, Bolte-2008, Dussaux-2010}). Although the latter approach is easier to be implemented in practice and appears to be more energy efficient, the issue of reducing the required current amplitude remains a high priority.
\par
An original solution could be to use pure spin currents, which can lead to the generation of large spin torques acting on magnetization. Such torques, which can be due to the interfacial Rashba effect but more often are the result of the bulk spin Hall effect, are considered for the ferromagnetic metal layer with asymmetric Pt and $\mathrm{AlO_x}$ interfaces \cite{Miron-2010} as well as for a graphene/Pt interface \cite{Shikin-2014}. Also such torques are observed in the case of ferromagnetic/Pt \cite{Liu-2011} or ferromagnetic/Ta \cite{Liu-2012} bilayers, and more generally in systems with a ferromagnet in contact with a layer of large spin orbit coupling material ($\mathrm{Pt}$, $\mathrm{Ir}$, $\mathrm{W}$, $\mathrm{Ta}$, $\mathrm{Pd}$ etc.). Using these torques, it is possible to excite domain wall motion \cite{Obata-2008, Khvalkovskiy-2013}, to prevent Walker breakdown \cite{Miron-2011}, to switch the magnetization of a ferromagnetic disk \cite{Miron-2011-2, Wang-2012, Liu-2012} and also to excite magnetization oscillations \cite{Demidov-2012, Liu-2013}. In this context, an important question is to find new materials with properties favourable the generation of larger spin currents. 
\par
Nowadays, bismuth selenide and bismuth telluride materials with giant spin-orbit interaction (SOI) attract a lot of attention mainly in the contest of topological insulators (TI). 
The possibilities of magnetization switching \cite{Roy-2015} and enhancement of the Walker breakdown \cite{Linder-2014} by TI were demonstrated theoretically. In 2014, the bias current induced spin polarization in Bi$_2$Se$_3$ was observed experimentally \cite{Li-2014}. Moreover, the magnetization switching through giant spin-orbit torque induced by an in-plane current in a chromium-doped TI bilayer heterostructure was demonstrated experimentally at low temperature \cite{Fan-2014}. In \cite{Wang-2015} spin-orbit torques in Bi$_2$Se$_3$/Co$_{40}$Fe$_{40}$B$_{20}$ devices were evaluated at different temperatures using ferromagnetic resonance measurements. It was demonstrated that the magnitudes of the spin-orbit torques decrease be factor of ten with increase of the temperature from 50 $K$ to 300 $K$, 
which indicates that origin of the observed spin-orbit torques in this system are topological surface states (TSS).  Most of these appealing TSS-based effects, however, were observed at low temperatures and for very pure materials. Considerably "dirty" bismuth selenides and bismuth tellurides, which can be or doped either have many structure defects, still have a giant SOI, and thus are very promising for spintronics. In this case the Fermi level is located in the conduction band, therefore, because of a very small statistical weight of TSS, the spin-orbit effects are determined mainly by the band states. Recent experimental works demonstrate that charge current flowing at the surface of a thin film of Bi$_2$Se$_3$ at room temperature can exert a spin-transfer torque on an adjacent ferromagnetic permalloy (Py) film, with the strength greater than of any other sources of spin-transfer torque measured thus far \cite{Mellnik-2014, Baker-2015}. Moreover, in case of doped material (like BiTeI) it was demonstrated experientially that although it ceases to be an insulator and demonstrates bulk conductance, it still possess strong spin-orbit coupling \cite{Ishizaka-2011}, which expected to be enough to induce the magnetization dynamics. 
\par
\begin{figure}[h!]
\centering
\includegraphics[width=0.45\textwidth]{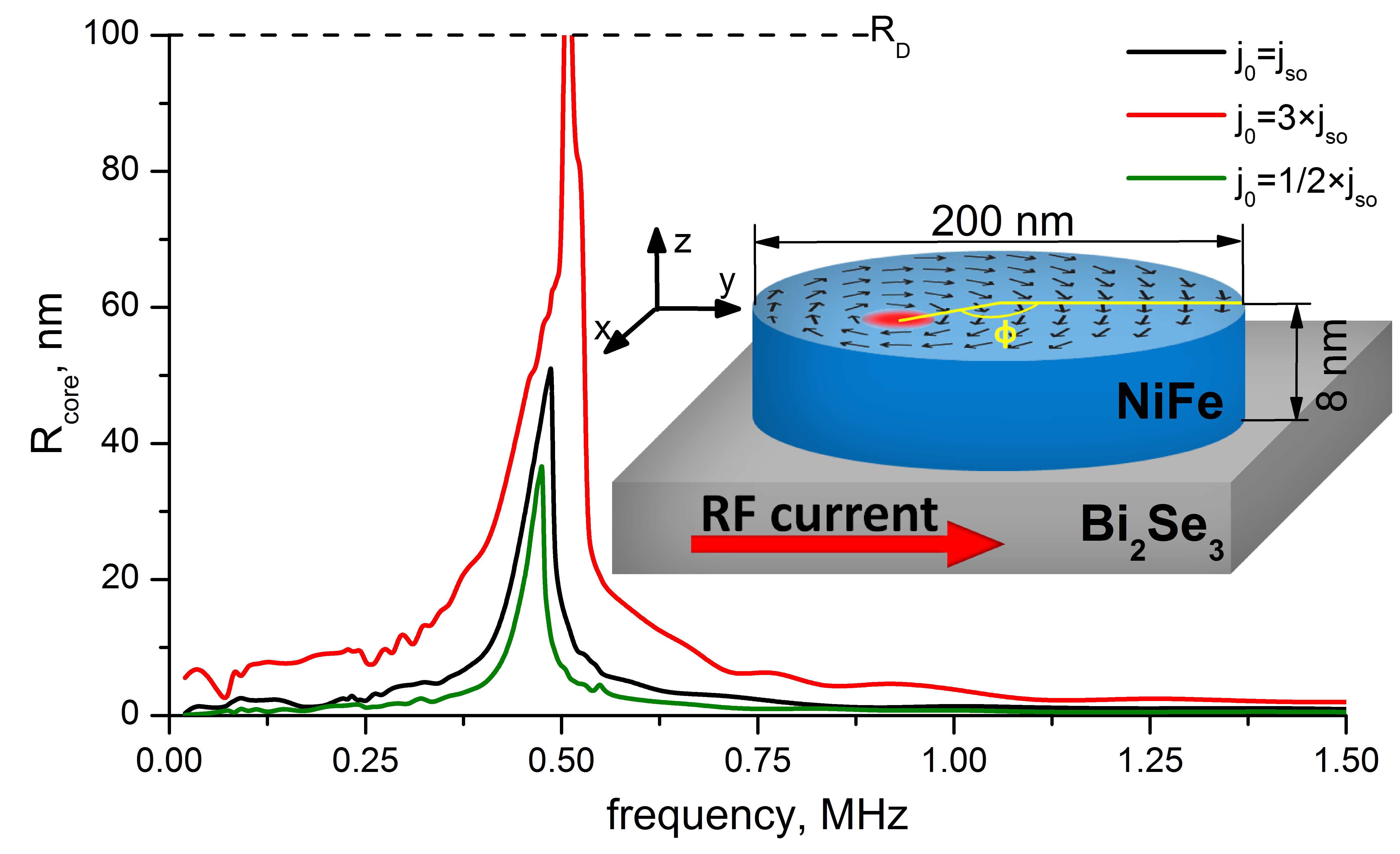}
\caption{(Color online) Vortex core orbit $R_{core}$ as a function of frequency of the input RF signal for $j_0=j_{so}$ (black line), $j_0=1/2\times j_{so}$ (green line) and $j_0=3\times j_{so}$ (red line), where $j_{so}$ is the characteristic
current density for the considered system. Inset: the system under investigation from afar. The black arrows represent magnetization direction, while the red region represents the vortex core with positive $m_z$. \label{fig:1}}
\end{figure}

This is one of our objectives: to evaluate the feasibility and effectiveness to excite gyrotropic vortex motion using torque originating at the Bi$_2$Se$_3$/Py interface. The studied system (see inset in Fig.~\ref{fig:1}) is represented by a $\mathrm{Ni_{81}Fe_{19}}$ disk with radius $R_D=100~nm$ and vortex magnetization distribution on top of a Bi$_2$Se$_3$ film. Such a system is very attractive for direct observations of magnetization dynamics by optical means since the ferromagnetic layer is open at the top in contrast to conventional STNOs or spin-diode effect which require a capping electrode \cite{Demidov-2012}. First, we demonstrate by micromagnetic modelling that although the torque associated with the direct in-plane current leads only to a finite vortex position shift, the ones associated with radio-frequency (rf) in-plane current along the interface can efficiently excite large rotation of the vortex core. We consider the dependence of this excitation on the frequency and the magnitude of the rf current. The analytical description of this phenomenon is proposed.
\par
Here we study the magnetization dynamics in the nanodisk under the action of an rf in-plane current. The disk thickness is $d=8~nm$. The magnetization dynamics in the NiFe disk is described by the Landau-Lifshitz-Gilbert (LLG) equation with an additional term responsible for the spin-transfer torque caused by TI:
\begin{equation}
\label{eq:LLG}
\dot{\mathbf{M}}=-\gamma\mathbf{M}\times\mathbf{H}_{eff}+\frac{\alpha}{M_S}(\mathbf{M}\times\dot{\mathbf{M}})+\mathbf{T}_{STT},
\end{equation}
where $\mathbf{M}$ is the magnetization vector, $\gamma$ is the gyromagnetic ratio, $\alpha$ is the Gilbert damping constant, $M_S$ is the saturation magnetization and $H_{eff}$ is the effective field consisting of the magnetostatic field, the exchange field and the anisotropy field. The spin-transfer torque $\mathbf{T}_{STT}$ can generally be represented by two components: a parallel (in-plane) torque $\mathbf{T}_{\parallel}=\gamma\tau_{\parallel}\mathbf{m}\times(\mathbf{e}_x\times\mathbf{m})$ and a perpendicular torque $\mathbf{T}_{\perp}=\gamma\tau_{\perp}(\mathbf{m}\times\mathbf{e}_x)$, where $\mathbf{e}_x$ is a unit in-plane vector perpendicular to current direction and $\mathbf{m}$ is a unit vector along the magnetization. In materials such as Bi$_2$Se$_3$, a giant Rashba interaction is commonly present. This interaction causes significant spin accumulation at the surface \cite{Li-2014}, that can interact directly by exchange with the magnetic moment of the adjacent NiFe disk. The diffusive spin current inside the adjacent ferromagnet (similar to the spin Hall effect) can create a parallel torque, and the combined actions of the Rashba effect and the exchange interaction can produce a perpendicular one. The microscopic details of these effect in studied system are still not completely clear, but the first experimental evidence of the existence of both torques has already been obtained \cite{Mellnik-2014}. In this case, the torque densities can be expressed as $\tau^{so}=-\operatorname{div}\hat{\mathbf{j}}_S$, where $\mathbf{j}_S$ is a spin current. Substituting the spin torque ratio $\theta=(j_S/j)(2e/\hbar)$, one can obtain torque density $\tau_{i}^{so}=\hbar\theta_{i}j/2ed$ (with $i=\parallel ,\perp$), where $j$ is the current density and $e>0$ is the charge of the electron.
\par
It needs no mention that the spin-transfer torque $\mathbf{T}_{STT}$ in general case may have more complex form than the one considered here (see Ref.~\cite{Garello-2013}). However for the magnetic vortex, which is formed by the in-plane curling magnetization and a tiny out-of-plane core, the dynamical properties are defined solely by the in-plane volume. Consequently in our case we can consider only the torques action on the in-plane component of the magnetization, without any loss of generality. In other words $\mathbf{m}(\mathbf{r})=(\cos\chi(\mathbf{r}), \sin\chi(\mathbf{r}), 0)^T$, where $\mathbf{m}(\mathbf{r})$ is a unit vector along the magnetization and $\chi(\mathbf{r})$ is the angle between X-axis and magnetization direction. Using it, general spin-transfer torque form from Ref.~\cite{Garello-2013} is reduced to $\mathbf{T}_{\perp}=T_{\perp}(\mathbf{m}\times\mathbf{e}_x)$ and $\mathbf{T}_{\parallel}=T_{\parallel}\mathbf{m}\times(\mathbf{e}_x\times\mathbf{m})$ respectively, when one consider the vortex-based system.
\par
To investigate the possibility of vortex core excitation by the current flowing along the interface Bi$_2$Se$_3$/Py, we have performed a series of simulations using our micromagnetic finite-difference code SpinPM based on the fourth-order Runge-Kutta method with an adaptive timestep control for the time integration and a mesh size $2\times2\times10~nm^3$. The larger mesh size in the z-direction has been chosen because the modelling the magnetization almost does not change with the z-axis. To focus on TI action, we have not included the Oersted field in our simulations, because its value is not enough for vortex core expulsion and even for significant excitation. The Py magnetic parameters used in the modelling are: $M_S=800~emu/cm^3$, the exchange constant is $A=1.3\times10^{-6}~erg/cm$, $\alpha=0.01$ and the bulk anisotropy is neglected.
\par
\begin{figure*}[t!]
\centering
\includegraphics[width=0.95\textwidth]{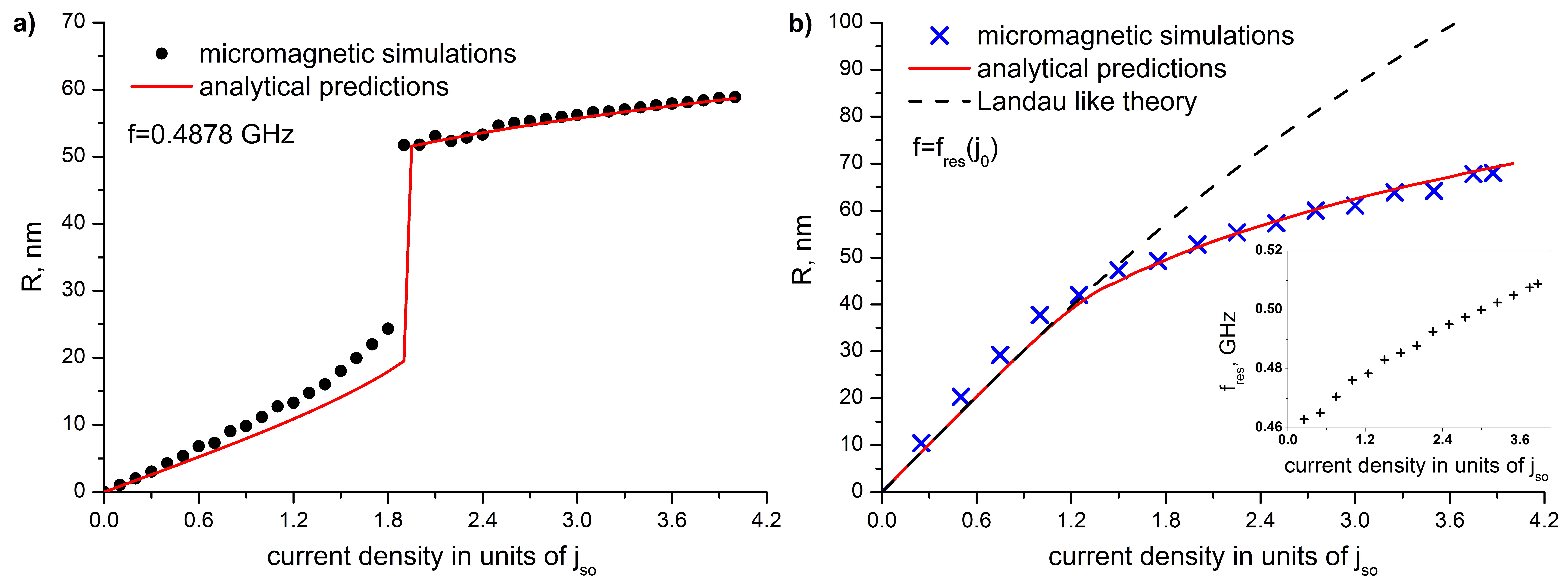}
\caption{(Color online) Vortex core stationary orbit versus amplitude $j_0$ of the rf current density. a) The orbit obtained by the micromagnetic modelling (black dots) and by numerical solution of Eq.\eqref{eq:sol1},\eqref{eq:sol2} (red line) in case of constant frequency, which is resonant for the average value of $j_0=j_{so}$. b) The orbit obtained by the micromagnetic modelling (blue crosses), by solutions of Thiele equations Eq.\eqref{eq:sol1},\eqref{eq:sol2} (red line), and by Landau like theory solution of Eq.\eqref{eq:j} (dash black line) in case, when the frequency is adjusted for every current amplitude $j_0$. The quantity $j_{so}$ here is chosen in order to introduce the figure in arbitrary units. Inset: the dependence of the resonant frequency of excitation on the current amplitude $j_0$. \label{fig:2}}
\end{figure*}
We find in micromagnetic modelling that the in-plane direct current cannot excite vortex gyration, but only slightly shifts the vortex core from the equilibrium position. Thus, subsequently, we will focus on the influence of rf currents. According to the micromagnetic modelling, after the application of an in-plane rf current with a density of $j=j_0\sin(\omega t)$, the vortex starts gyrotropic motion after a transitional period. Then it reaches a stationary orbit due to the action of the spin torque created at the TI/Py interface. The dependences of the vortex core orbit $R_{core}$ on the frequency of the input RF signal for three different alternating current density amplitudes $j_0$, are shown in Fig.~\ref{fig:1} with the vortex core being resonantly excited. At the same time, an increase of the current amplitude $j_0$ leads to an increase in the stationary orbit of the vortex core. Moreover, if the value $j_0$ becomes higher than critical value $j_0^{cr}$, the vortex core gets ejected out of the dot (corresponds to the region on Fig.~\ref{fig:1} where the red line overtakes the $R_D$ value). It is also worth mentioning that the excitation has nonlinear behaviour for high current values, similar to the Ref.~\cite{Drews-2012}. For further analysis, let us introduce the dimensionless currents ratio $j/j_{so}$, where $j_{so}$ is the characteristic current density for the considered system, and it is chosen in order to introduce the figure in arbitrary units.
\par
For analytical insight into the vortex excitation we analyze the Thiele equation which can be deduced from Eq.\eqref{eq:LLG}:
\begin{equation}
\label{eq:Thiele}
G(\mathbf{e}_z\times\dot{\mathbf{R}})=k(\mathbf{R})\mathbf{R}+D\dot{\mathbf{R}}-\mathbf{F}_{ST}
\end{equation}
where $\mathbf{R}$ is the vortex core position, $G=-2\pi p M_S h/\gamma$ is the gyroconstant, $p$ is the core polarity, and $\mathbf{e}_z$ is the unit vector along the z-axis. The confining force is given with $k(\mathbf{R})=\omega_0G(1+a\mathbf{R}^2/R_D^2)$ where the gyrotropic frequency is $\omega_0=\frac{20}{9}\gamma M_S h/R_D$ and $a\approx0.25$. The damping coefficient is $D=\alpha G(\frac{1}{2}\ln(\frac{R_D}{2l_e})+\frac{3}{8})$ \cite{Khvalkovskiy-2009} where $l_e=\sqrt{\frac{A}{2\pi M_S^2}}$. As mentioned before, the spin-transfer force $\mathbf{F}_{ST}$ consist of two contributions: a perpendicular torque and a parallel torque. The first one modifies the energy of the vortex with the additional term $E_{\perp}=-\delta(\mathbf{H_{eff}}\times\mathbf{e}_z)\mathbf{R}$, therefore it can be represented as $\mathbf{F}_{\perp}=\delta(\mathbf{e}_z\times\mathbf{H_{eff}})$, where $\mathbf{H_{eff}}$ is the effective field corresponding to the perpendicular torque. This formula is already in use in the case of external magnetic field \cite{Guslienko-2006}. Because the perpendicular torque has exactly the same symmetry, hence it can be used here with the effective field: $\mathbf{H_{eff}}=\tau_{\perp}^{so}\sin(\omega t)\mathbf{e}_x$. Micromagnetic modelling gives $\delta=5.6~\frac{emu/cm^3}{nm}$. Using the Feldtkeller ansatz \cite{Feldtkeller-1965} for describing the vortex configuration, the in-plane torque contribution can be represented by $\mathbf{F}_{\parallel}=\beta G(\cos(\phi)\mathbf{e}_x-\sin(\phi)\mathbf{e}_y)$, where $\beta=\frac{\pi^{3/2}}{8}\rho_cV\gamma\tau_{\parallel}^{so}\sin(\omega t)$, $\rho_c$ is the vortex core width, $V$ is the vorticity and $\phi$ is the polar angle of the core. Using this, Eq.\eqref{eq:Thiele} can be represented in polar coordinates $(R,\phi)$ in the following form:
\begin{align}
\label{eq:sol1}
& \frac{\dot{R}}{R}=\frac{D}{G}\dot{\phi}-\frac{\delta}{G}\frac{H(t)}{R}\cos(\phi)+\frac{\beta}{R}\sin(2\phi) \\ \label{eq:sol2}
& \dot{\phi}=-\frac{k(R)}{G}-\frac{D}{G}\frac{\dot{R}}{R}+\frac{\delta}{G}\frac{H(t)}{R}\sin(\phi)+\frac{\beta}{R}\cos(2\phi)
\end{align}
where $H(t)=\tau_{\perp}^{so}\sin(\omega t)$.
\par
The dependence of the stationary orbit of the vortex core on the amplitude $j_0$ of the rf current density is represented in Fig.~\ref{fig:2}. The resonant frequency of excitation increases with the increase of the current amplitude $j_0$ (see inset in Fig.~\ref{fig:2}.b). Hence, we consider two different dependencies of the orbit on the amplitude $j_0$: with fixed frequency, which is resonant for the average value of $j_0$ (see Fig.~\ref{fig:2}.a), and with adjusting of the frequency for every $j_0$ (see Fig.~\ref{fig:2}.b). The first regime is more easily implemented, while the second one demonstrates greater efficiency of excitation, because it always deals with resonant frequency. The dependencies of the orbit $R_{core}$ on the amplitude $j_0$ for constant frequency, which is resonant for $j_0=j_{so}$, obtained by the micromagnetic modelling (black dots) and by numerical solution of Eq.\eqref{eq:sol1},\eqref{eq:sol2} (red line), are shown on Fig.~\ref{fig:2}.a. The jump of the function, corresponding to a sharp change of orbit, arises from the fact that at the left side of the current $j_{so}$, which is resonant for the selected frequency, the excitation efficiency is reduced due to the frequency mismatch. At the same time, this frequency mismatch is compensated by increasing of the amplitude at the right side of the resonance current.
\par
In Fig.~\ref{fig:2}.b, the dependence of the orbit on the amplitude $j_0$ for the frequency is adjusted for every $j_0$ obtained by the micromagnetic modelling (blue crosses) and by the numerical solution of Eq.\eqref{eq:sol1},\eqref{eq:sol2} (red line). It should be noted that the orbit in case of fixed frequency is always less than in case of adjusted frequency, except for the jump point, where they are equal. As follows from these results, the analytical solution demonstrates good coincidence with the micromagnetic predictions.
\par
Both the numerical simulations and analytical Thiele modelling demonstrate that the instant frequency of vortex core rotation oscillates during the period with the deviation amplitude less than $4\%$, while the mean frequency is equal to the frequency of the alternating current. Taking this into account, we can integrate Eq.\eqref{eq:sol1},\eqref{eq:sol2} over the period under the approximation that the frequency is constant. In this case, some oscillating terms vanish, while others return constant values after integration, and the Eq.\eqref{eq:sol1} then takes the form:
\begin{equation}
\label{eq:rho}
\dot{\rho}=-\tilde{D}\omega_0\rho-\tilde{D}\omega_0 a\rho^3+h
\end{equation}
where $\rho=R/R_D$, $\tilde{D}=D/G$ and $h=\delta\tau_{\perp}^{so}/2GR_D$. Let us define function $\mathcal{F}$ as $\dot{\rho}=-\partial \mathcal{F}/\partial\rho$. Then using Eq.\eqref{eq:rho} function $\mathcal{F}$ can be represented in the form:
\begin{equation}
\label{eq:F}
\mathcal{F}=-h\rho+\tilde{D}\omega_0\frac{\rho^2}{2}+\tilde{D}\omega_0 a\frac{\rho^4}{4}
\end{equation}
This function is similar to a typical free energy functional in Landau theory, if we consider $\rho$ as an order parameter and $h$ as an external field conjugate to the order parameter. In this notation, Eq.\eqref{eq:rho} is equivalent to the Landau-Khalatnikov equation. The stationary orbit due to Eq.\eqref{eq:rho} can be found by solving $\partial \mathcal{F}/\partial\rho=0$. This condition coincides with energy functional minimum condition. These analogies are valid not only for rf torques produced by TI, but also for all possible oscillating torques with the same symmetry (for example produced by the Rashba interaction or by the external magnetic field or by the Oersted field). In this case, the dependence of the stationary orbit of the vortex core on the amplitude $j_0$ of the rf current can be represented in the form:
\begin{equation}
\label{eq:j}
h/\tilde{D}\omega_0=\rho+a\rho^3
\end{equation}
The solution of Eq.\eqref{eq:j} is shown on Fig.~\ref{fig:2}.b (dash black line). It demonstrates good agreement with Thiele modelling and the numerical solution of Eq.\eqref{eq:sol1},\eqref{eq:sol2} until the orbits about $R_D/2=50~nm$. For orbits greater than this value, the larger oscillations of the frequency during the period appears, which makes our approximation no more valid.
\par
Experimental observation of spin torque ratio for Bi$_2$Se$_3$ with an $8~nm$ permalloy layer atop \cite{Mellnik-2014} gives $\theta_{\parallel}\ge 3.5$. It is worthy to note that this value already takes into account shunting by the Permalloy layer. To avoid shunting problems completely and therefore to obtain the maximum possible efficiency, the permalloy layer can be changed by the insulating ferromagnetic layer, as it proposed in Ref.~\cite{Mellnik-2014}. At the same time, the latest results of the same group \footnote{D.C. Ralph, APS March Meeting 2015 in San Antonio, Texas, March 2-6} give lower estimations for the considered system, which are about $\theta_{\parallel}\approx\theta_{\perp}\approx 1$. Using these values of spin-torque ratio let us consider the magnitudes of the torques with field-like symmetry, which is responsible for vortex gyration considered in this work, in different cases. In case of conventional perpendicular current injection (CPP) the magnitude $\tau^{cpp}_{\perp}=\xi\frac{\hbar PJ}{2ed}$, where $\xi=0.1-0.4$ and $P$ is the current polarization for a Magnetic Tunnel Junction with typical values $P=0.3-0.4$. In case of excitation by spin-orbit effects the magnitude $\tau^{so}_{\perp}=\frac{\hbar \theta_{\perp} J}{2ed}$. Therefore, the comparison between CPP and spin-orbit excitation gives $\tau^{so}_{\perp}/\tau^{cpp}_{\perp}=\frac{\theta_{\perp}}{\xi P}\approx 10$ in case of Bi$_2$Se$_3$. These estimations give us the following value of $j_{so}=1.5\times10^6~A/cm^2$. It means that in Permalloy/Bi$_2$Se$_3$-like heterostructures the created torque can be significantly more effectively than in case of perpendicular current injection, which is mostly used nowadays.
\par
In conclusion, we have demonstrated the possibility of vortex oscillations excitation by the torques caused by Bi$_2$Se$_3$-like structures. Using micromagnetic modelling, we prove that in-plane rf current along the interface of the Bi$_2$Se$_3$/Py heterostructure can excite vortex core rotation with a required current density several times lower than previously observed. We consider the dependence of vortex excitation on the frequency and the magnitude of the rf current both by micromagnetic modelling and by analytical theory. An analytical description of the system under consideration is proposed. The analogy between the used approach and the Landau theory of phase transitions is demonstrated. On this basis, the vortex excitation by the RF current through the the interface of the Bi$_2$Se$_3$/Py heterostructures becomes a very promising option for practical applications in spin-transfer nanooscillators, spin-torque diodes and so forth. This new fundamental mechanism of vortex excitation and peculiarities of its dynamics remain valid not only for considered material, but also for any alternative sources of the spin-currents with in-plane spin polarization, like Bi-based materials and other materials with strong spin-orbit coupling. However, other materials likely to provide smaller magnitude of the effect than Bi$_2$Se$_3$.
\par
We acknowledge Ching-Ray Chang, Romain Lebrun and Alex Jenkins for the helpful discussions. Financial support by the RFBR Grants No. 12-02-01187, No. 14-02-31781 and 50 Labs Initiative of Moscow Institute of Physics and Technology is acknowledged.
\providecommand{\noopsort}[1]{}\providecommand{\singleletter}[1]{#1}%
\end{document}